\documentclass[aps,preprint]{revtex4-2}
\usepackage{amsmath}
\usepackage{bm}
\usepackage{graphicx}
\usepackage{amssymb}
\usepackage{xcolor}

\begin{document}

\title{Confinement-Induced Nonlocality and Casimir Force in Transdimensional Systems}

\author{Igor V. Bondarev}\email[Corresponding author: ]{ibondarev@nccu.edu}
\author{Michael D. Pugh}
\affiliation{Department of Mathematics \& Physics, North Carolina Central University, Durham, NC 27707, USA}

\author{Pablo Rodriguez-Lopez}
\affiliation{\'{A}rea de Electromagnetismo and Grupo Interdisciplinar de Sistemas Complejos (GISC), Universidad Rey Juan Carlos, 28933 M\'{o}stoles, Madrid, SPAIN}
\affiliation{Laboratoire Charles Coulomb (L2C), UMR 5221 CNRS-University of Montpellier, F-34095 Montpellier, FRANCE}

\author{Lilia M. Woods}
\affiliation{Department of Physics, University of South Florida, Tampa, FL 33620, USA}

\author{Mauro Antezza}
\affiliation{Laboratoire Charles Coulomb (L2C), UMR 5221 CNRS-University of Montpellier, F-34095 Montpellier, FRANCE}
\affiliation{Institut Universitaire de France, 1 rue Descartes, F-75231 Paris Cedex 05, FRANCE}

\begin{abstract}
We study within the framework of the Lifshitz theory the long-range Casimir force for in-plane isotropic and anisotropic free-standing transdimensional material slabs. In the former case, we show that the confinement-induced nonlocality not only weakens the attraction of ultrathin slabs but also changes the distance dependence of the material-dependent correction to the Casimir force to go as $\sim\!1/\sqrt{l}$ contrary to the $\sim\!1/l$ dependence of that of the local Lifshitz force. In the latter case, we use closely packed array of parallel aligned single-wall carbon nanotubes in a dielectric layer of finite thickness to demonstrate strong orientational anisotropy and crossover behavior for the inter-slab attractive force in addition to its reduction with decreasing slab thickness. We give physical insight as to why such a pair of ultrathin slabs prefers to stick together in the perpendicularly oriented manner, rather than in the parallel relative orientation as one would customarily expect.
\end{abstract}

\maketitle

\newpage

\section{Introduction}\label{Sec:1}

Modern fabrication techniques have appreciably improved the quality of thin films, making it possible to produce ultrathin films of precisely controlled thickness down to a few monolayers~\cite{thingold,Shah17,Shah18,ZhelNatCom18,MariaNL19,javierOptica19,GarciaAbajoACS19,TunableGold,SnokeBond21,NL22TiN}.~Such ultrathin films, often referred to as transdimensional (TD) quantum materials or TD metasurfaces (MSs)~\cite{BoltShalACS19,NL22TiN,Shah2020}, offer high tailorability of their electronic and optical properties not only by altering their chemical and electronic composition (stoichiometry, doping) but also by varying their thickness (the number of monolayers)~\cite{Manjavacas22,BondADP22,CommPhys-bond,magnons,CNArr21PRAppl,CNArr21JAP,BondPRR20,BondOMEX19,Koppens18,BondMRSC18,Brongersma17,BondOMEX17}. Whereas three-dimensional (3D) bulk materials allow for higher free carrier concentration, and their two-dimensional (2D) counterparts such as graphene and monolayer transition metal dichalcogenides provide the strong confinement of exciton-polariton and plasmon modes~\citep{Basov2014,Mak2016,Xia2014}, the advantages of both of these extremes can be merged by using TD quantum systems~\cite{BoltShalACS19}. For such systems, quite generally, the vertical quantum confinement leads to an effective dimensionality reduction from 3D to 2D while still retaining the thickness as a parameter to represent the vertical size~\cite{KRR,KRK,Kleef}.

Since the TD regime is situated in-between the 3D and 2D dimensionalities, quantum-confined TD materials make it possible to probe fundamental properties of light-matter interactions as they evolve from a single atomic layer to a larger number of layers approaching the bulk material properties~\cite{BondPRR20,BondADP22}.~Ultrathin films of metals, doped semiconductors or polar materials with thickness of only a few monolayers can support plasmon, exciton and phonon-polariton modes~\cite{Shah18,ZhelNatCom18,MariaNL19,javierOptica19,GarciaAbajoACS19,TunableGold,SnokeBond21,NL22TiN}. Due to their localized plasmon modes~\cite{BondOMEX17,BondOMEX19} and associated thickness-dependent photonic density of states~\cite{Manjavacas22,BondPRR20}, they can provide controlled light confinement and tunable light-matter coupling which makes them distinctly different from conventional thin films. Fundamental properties of the TD systems originate from the quantum-confined electromagnetic (EM) dispersion of their eigen excitation modes and cannot be inferred from those of 2D systems or 3D materials with boundary conditions imposed on their top and bottom interfaces~\cite{BondPRR20}. Vertical confinement makes these modes nonlocal and so distinct from those of conventional thin films. Confinement-induced nonlocality is the remarkable intrinsic property of the in-plane EM response of TD systems~\cite{BondOMEX17}. It is this nonlocality that enables a variety of new quantum phenomena in ultrathin films, including thickness-controlled plasma frequency red shift~\cite{NL22TiN,PCCP22}, low-temperature plasma frequency dropoff~\cite{Lavrinenko19}, plasmon mode degeneracy lifting and spontaneous emission enhancement~\cite{BondPRR20}, directional negative refractivity~\cite{CNArr21PRAppl}, absorption-transmission switching under controlled exciton-plasmon coupling~\cite{CNArr21JAP}, a series of quantum-optical~\cite{BondADP22}, magneto-optical~\cite{BondMRSC18} and radiative heat transfer effects~\cite{BiehsBond2022,BondBiehsShen} as well as quantum electronic transitions that are normally forbidden~\cite{Rivera}.

\begin{figure}[t]
\includegraphics[width=1.0\linewidth]{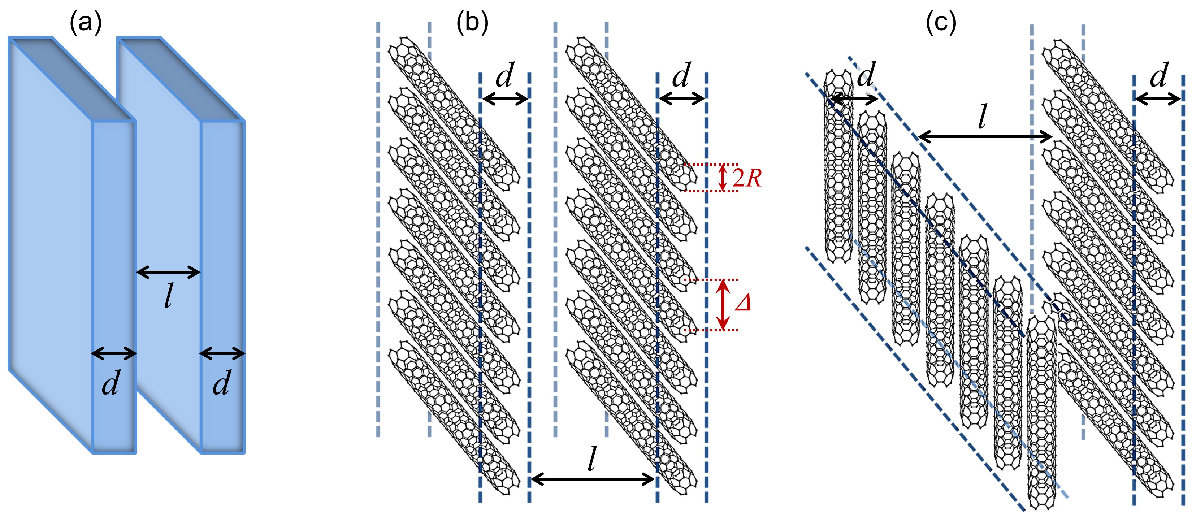}
\caption{Finite-thickness parallel material slabs studied in this work. (a)~Two identical in-plane isotropic slabs. (b),(c)~Two relative orientations for in-plane anisotropic slabs made of parallel-aligned periodic SWCN arrays. Monolayers shown are just for a sketch. The actual slabs we study are composed of a varied number of densely-packed monolayers. See text for details.}
\label{fig1}
\end{figure}

The confinement-induced nonlocal EM response comes from the Keldysh-Rytova (KR) pairwise interaction potential~\cite{BondOMEX17,BondOMEX19}. The KR interaction potential takes into account the vertical electron (and/or hole) confinement due to the presence of substrate and superstrate with dielectric permittivities much less (e.g., air) than the in-plane dielectric permittivity of the film itself~\cite{KRK,KRR}. For ultrathin films such a potential is much stronger than the Coulomb interaction potential. It transitions into the electrostatic Coulomb potential with film thickness increase, which is why the thickness can be treated as a parameter to control the optical response of TD films. The nonlocal KR model of the EM response is unique in that it covers the entire range from atomically thin to conventional films of the order of a few optical wavelengths in thickness~\cite{NL22TiN,BondPRR20}. Previously~\cite{BiehsBond2022}, a comparative analysis of the nonlocal and standard local (Drude) EM response models reported significant differences in their predictions of far- and near-field heat transfer properties of the TD film systems. Very recently, the nonlocal EM response model was successfully tested experimentally~\cite{BondBiehsShen}. It is this nonlocal model that we use here for theoretical studies into a closely related phenomenon --- the Casimir effect in ultrathin TD systems. The Lifshitz theory we employ herein~\cite{Lifshitz,LandauLifshitz}, conveniently represents molecular attractive forces between solids in terms of their respective linear EM response functions. We calculate the long-range attractive force between two free-standing ultrathin TD material slabs as functions of their thickness and separation distance. The geometries considered are shown in Fig.~\ref{fig1}, where (a) is the case of two identical in-plane isotropic slabs, while (b) and (c) sketch the two relative orientations for in-plane anisotropic finite-thickness slabs made of densely-packed parallel-aligned periodic single-wall carbon nanotube (SWCN) arrays studied recently in Ref.~\cite{CNArr21PRAppl}. We show that the confinement-induced nonlocality of the EM response leads to significant reduction of the attractive force. In the in-plane anisotropic case, additionally, we predict strong orientational anisotropy and crossover behavior for the attractive force in the (b) and (c) orientations depending on the film thickness and SWCN diameter variation. The following sections describe our model, present our theoretical results, and conclude our work by summarizing its key findings.

\section{Casimir effect for a pair of parallel TD material slabs}

We start with the long-range attractive force expression of the Lifshitz theory for molecular attractive forces between solids at zero temperature~\cite{Lifshitz,LandauLifshitz}. In this case the separation distance between the slabs sketched in Fig.~\ref{fig1} is larger than the fundamental wavelength of their absorption spectra, whereby the force takes the following form~\cite{Lifshitz}
\begin{eqnarray}
F=\frac{\hbar c}{32\pi^2l^4}\int_{0}^{\infty}\!\!\!\!\!\!dx\!\!\int_1^\infty\!\!\!\!\!\!dp\,\frac{x^3}{p^2}\Big\{\Big[\frac{(s_1+p)(s_2+p)}{(s_1-p)(s_2-p)}e^x-1\Big]^{-1}
+\frac{(s_1+\varepsilon_1p)(s_2+\varepsilon_1p)}{(s_1-\varepsilon_1p)(s_2-\varepsilon_1p)}e^x-1\Big]^{-1}\Big\},\nonumber\\
\label{Lifshitz}\\
\varepsilon_{1,2}=\varepsilon_{1,2}\Big(i\frac{xc}{2pl}\Big),\;\;\;s_{1,2}=\sqrt{\varepsilon_{1,2}\Big(i\frac{xc}{2pl}\Big)-1+p^2}.\hskip3cm\nonumber
\end{eqnarray}
Here, $c$ is the speed of light, $l$ is the distance between the slabs, and $\varepsilon_{1,2}(\omega)$ are their respective linear EM response functions (dynamical dielectric permittivities) treated as functions of the complex variable $\omega\!=\!i\xi$ with $\xi\!=\!xc/(2pl)$.

For a pair of perfectly conducting metallic plates, in which case $\varepsilon_{1,2}\!=\!\infty$, Eq.~(\ref{Lifshitz}) can be integrated exactly to give the well known Casimir force~\cite{Casimir}
\begin{equation}
F_C=\frac{\hbar c}{l^4}\frac{\pi^2}{240}\,.
\label{Casimir}
\end{equation}
In terms of the Lifshitz theory, however, the Casimir force is just the zero-order expansion term of Eq.~(\ref{Lifshitz}). The next expansion term can be obtained if one uses the explicit form of the dynamical dielectric functions $\varepsilon_{1,2}(\omega)$ in the frequency range that contributes the most to the integral in Eq.~(\ref{Lifshitz}). It can be seen that this is the domain where $p\!\sim\!1$ and $x\!\sim\!1$, or $\omega/c\!\sim\!1/l\!\sim\!0$ since $x\!=\!2pl\xi/c$ and $\omega\!=\!i\xi$ as per the Lifshitz theory in the large separation limit~\cite{Lifshitz}. For normal metals commonly described by the standard low-frequency local (Drude) EM response function
\begin{equation}
\varepsilon(\omega)=\varepsilon_b-\frac{\omega_p^2}{\omega(\omega+i\delta)}\,,
\label{Drude}
\end{equation}
where $\epsilon_{b\,}$ is the constant background permittivity, $\delta$ stands for the damping constant,
\begin{equation}
\omega_p=\omega_p^{3D}=\sqrt{\frac{4\pi e^2N_{3D}}{m^\ast}}\,,
\label{DrudePlasma}
\end{equation}
$\omega_p^{3D}$ is the bulk plasma frequency, and other parameters have their usual meanings, the Casimir force in Eq.~(\ref{Casimir}) comes out of the Lifshitz theory as the main (zero-order) power series expansion term of the integrand in Eq.~(\ref{Lifshitz}) that one obtains in the $\varepsilon(\omega\!=\!0)$ limit. The correction to it can be obtained by taking into account the second term of Eq.~(\ref{Drude}), $\Delta\varepsilon\!=\!-\omega_p^2/\omega^2$ (damping neglected for simplicity), to the first nonvanishing order of the same integrand power series expansion under the condition $1/\!\sqrt{\Delta\varepsilon_{1,2}}\ll\!p\ll\!\sqrt{\Delta\varepsilon_{1,2}}$. For identical metals this gives~\cite{Lifshitz}
\begin{equation}
F_L=F_C\left(1-\frac{16c}{3\omega_pl}\right)
\label{Lifshitzforce}
\end{equation}
referred to as the Lifshitz force in what follows, where the second term in parentheses is the legitimate correction to the first at separations greater than that at which they are comparable. This term cannot be derived by the method of Ref.~\cite{Casimir} where the main term (\ref{Casimir}) was first obtained.

\subsection{In-plane isotropic TD systems}

For in-plane isotropic TD material films, due to the KR pairwise interaction potential of the charge carriers in it, the in-plane plasma oscillation frequency is given by the nonlocal expression as follows~\cite{BondOMEX17}
\begin{equation}
\omega_p(k)=\frac{\omega_p^{3D}}{\sqrt{1+1/(\tilde{\varepsilon}kd})}\,.
\label{omegapkd}
\end{equation}
Here, $\tilde{\varepsilon}\!=\!\varepsilon_b/(\varepsilon_1+\varepsilon_2)$ with $\varepsilon_{1,2}\;(<\!\varepsilon_b)$ being the film substrate and superstrate static permittivities, $\varepsilon_b$ is the in-plane dielectric permittivity of the film (contributed by both positive ion background and interband electronic transitions), $d$ is its thickness, and $k$ is the in-plane electron momentum absolute value. The low-energy in-plane EM response function of the film is still given by Eq.~(\ref{Drude}), but now with $\omega_p$ replaced by that of Eq.~(\ref{omegapkd}), which makes the in-plane dielectric response function $k$-dependent and so spatially dispersive, or nonlocal. This is the essence of the nonlocal EM response model for TD quantum materials~\cite{BondPRR20,BiehsBond2022}. With $d$ decreasing, it can be seen that $\omega_p(k)$ shifts to the red and Eq.~(\ref{omegapkd}) acquires the $\sqrt{k}$-type nonlocal spatial dispersion of 2D materials. As $d$ increases and becomes sufficiently large, Eq.~(\ref{omegapkd}) can be seen to gradually approach $\omega_p^{3D}$, the bulk material screened plasma frequency~(\ref{DrudePlasma}), and the EM response function (\ref{Drude}) takes the standard local Drude form.

The procedure described above to obtain the Lifshitz force (\ref{Lifshitzforce}) can be repeated for a pair of identical in-plane isotropic TD material slabs, sketched in Fig.~\ref{fig1}~(a), using the nonlocal EM response function of the slabs as given by Eqs.~(\ref{Drude}) and (\ref{omegapkd}). After having done the above mentioned substitution of variables and having also used the equality $\omega p/c\!=\!\sqrt{(\omega/c)^2-k^2}$ for the in-plane momentum $k$ to fulfill~\cite{Lifshitz,LandauLifshitz}, Eq.~(\ref{Lifshitz}) can be simplified to give the thickness dependent expression as follows
\begin{equation}
F=F_C\!\left[1-\frac{15c}{\pi^4\omega_p^{3D}l}\!\int_{0}^{\infty}\!\!\!\!\!\!dx\frac{x^4e^x}{\big(e^x-1\big)^{\!2}}\!\int_1^\infty\!\!\!\!\!\!dp\,\frac{p^2+1}{p^4}
\sqrt{1+\frac{2l}{\tilde{\varepsilon}d}\frac{p}{x\sqrt{p^2-1}}}\,\right].
\label{NLiso}
\end{equation}
In this equation the square root can be seen to tend to unity as $d\!\rightarrow\!\infty$. The remaining integral can be done analytically to give the Lifshitz force (\ref{Lifshitzforce}). In the opposite limit where $d$ becomes sufficiently small (see Sec.$\,$III$\;$C below for more precise definition of this term), the equation takes the following form
\begin{equation}
F=F_C\!\left[1-\frac{15\sqrt{2}c}{\pi^4\omega_p^{3D}\sqrt{\tilde{\varepsilon}dl}}\!\int_{0}^{\infty}\!\!\!\!\!\!dx
\frac{x^{7/2}e^x}{\big(e^x-1\big)^{\!2}}\!\int_1^\infty\!\!\!\!\!\!dp\,\frac{p^2+1}{p^{7/2}\sqrt[4]{p^2-1}}\right]
=F_C\!\left(1-\frac{4.79c}{\omega_p^{3D}\sqrt{\tilde{\varepsilon}dl}}\right).
\label{NLisothin}
\end{equation}
Here, the correction term in parentheses can be seen to go with inter-slab separation distance as $1/\sqrt{l}$, slower than the $1/l$ dependence of the correction term in the Lifshitz force (\ref{Lifshitzforce}) but with material and thickness dependent coefficient, which is the manifestation of the confinement-induced EM response nonlocality.

\subsection{In-plane anisotropic TD systems}

As an example of an in-plane anisotropic TD system we focus on a closely packed array of periodically aligned SWCNs of radius $R$, with translational unit $\Delta$, embedded in a dielectric layer of thickness $d$ as shown in Fig.~\ref{fig1}~(b) and (c). Material systems like that are currently in the process of intensive experimental development~\cite{Kono2016,Kono2020}, with a great potential to become the next generation advanced flexible platform for multifunctional metasurfaces and nonlinear optical devices with adjustable characteristics on demand~\cite{Brady2016,Falk2018,FalkFanNL2019,Kono2019,Naik2019,Liu2020,Fan2020,Fan2022}.

Collective EM response of a TD material slab made of periodically aligned SWCN arrays was recently studied theoretically~\cite{BondOMEX19,CNArr21PRAppl}. Being contributed by both plasmons and excitons (corresponding to intra- and interband transitions in the infrared and optical spectral regions, respectively), it was shown to be strongly anisotropically nonlocal due to the cylindrical spatial anisotropy, periodic in-plane transverse inhomogeneity, and vertical quantum confinement of the system. In the direction perpendicular to the SWCN alignment the in-plane response is a constant dielectric permittivity $\varepsilon_b$ effectively. In the SWCN alignment direction it is the complex-valued momentum-dependent (nonlocal) dynamical function given by Eq.~(\ref{Drude}) in the low-energy region of interest here~\cite{CNArr21PRAppl}, but now with~\cite{BondOMEX19}
\begin{equation}
\omega_p(q)=\omega_p^{3D}\sqrt{\frac{2qRI_0(qR)K_0(qR)}{1+1/(q\tilde{\varepsilon}d)}\,}
\label{plasmaFy}
\end{equation}
representing the intraband plasma oscillation frequency for a finite-thickness, cylindrically anisotropic, periodically aligned (metallic or semiconducting) SWCN array. Here, $\omega_p^{3D}$ is given by Eq.~(\ref{DrudePlasma}) with $N_{\rm 3D}\!=\!N_{\rm 2D}/d$, $q$ stands for the absolute value of the electron momentum along the SWCN alignment direction and makes the EM response of the slab unidirectionally nonlocal, $I_0$ and $K_0$ are the zeroth-order modified cylindrical Bessel functions responsible for the correct normalization of the electron density distribution over cylindrical surfaces, to give for $R\!\rightarrow\!\infty$, whereby $qRI_0(qR)K_0(qR)\!\rightarrow\!1/2$, the isotropic TD film plasma frequency of Eq.~(\ref{omegapkd}) studied previously in Refs.~\cite{BondPRR20,BiehsBond2022} and tested experimentally for TiN films of varied thickness both at room~\cite{NL22TiN} and at cryogenic temperatures~\cite{Lavrinenko19}. This EM response function features no bandstructure of individual SWCNs as it is contributed by intraband transitions alone, in which case their intrinsic properties are only determined by the carbon-carbon overlap integral (accounted for in $N_{2D}$) and by the surface curvature represented by $R$ here.

By symmetry, there are two minimum-energy relative orientations possible for a pair of identical space-separated finite-thickness SWCN slabs. They are shown in Fig.~\ref{fig1}~(b) and (c). Their respective attractive forces $F_\parallel$ and $F_{\perp}$ can be obtained from the Lifshitz formula of Eq.~(\ref{Lifshitz}) by noticing that the two terms under the integral sign there represent the inter-slab virtual $s$- and $p$-polarized photon exchange, respectively. Since (i)~we only have one direction for plasmon propagation on each of the slabs -- the one defined by $q$-vector of Eq.~(\ref{plasmaFy}) which is along the SWCN alignment direction, (ii)~plasmons can only be excited by $p$-polarized photons, and (iii)~each of the slabs is a dielectric in the direction perpendicular to the SWCN alignment direction, the two attractive forces in (b) and (c) orientations of identical slabs in Fig.~\ref{fig1} come out of Eq.~(\ref{Lifshitz}) in the form as follows
\begin{eqnarray}
F_{\parallel}=\frac{\hbar c}{32\pi^2l^4}\int_{0}^{\infty}\!\!\!\!\!dx\!\int_1^\infty\!\!\!\!\!dp\,\frac{x^3}{p^2}\Big\{\Big(\varphi^2e^x-1\Big)^{\!-1}\!+\Big[\Big(\frac{s+\varepsilon p}{s-\varepsilon p}\Big)^{\!2}e^x-1\Big]^{\!-1}\Big\},\hskip0.3cm\nonumber\\
F_{\perp}=\frac{\hbar c}{32\pi^2l^4}\int_{0}^{\infty}\!\!\!\!\!dx\!\int_1^\infty\!\!\!\!\!dp\,\frac{x^3}{p^2}\Big[\Big(\varphi\frac{s+p}{s-p}e^x-1\Big)^{\!-1}\!+\Big(\psi\frac{s+\varepsilon p}{s-\varepsilon p}e^x-1\Big)^{\!-1}\Big],\nonumber\\[-0.15cm]
\label{Fparper}\\[-0.15cm]
\varepsilon=\varepsilon\Big(i\frac{xc}{2pl}\Big),\;\;\;s=\sqrt{\varepsilon\Big(i\frac{xc}{2pl}\Big)-1+p^2},\hskip2.7cm\nonumber\\
\varphi=\frac{\sqrt{\varepsilon_b-1+p^2}+p}{\sqrt{\varepsilon_b-1+p^2}-p},\;\;\;\psi=\frac{\sqrt{\varepsilon_b-1+p^2}+\varepsilon_bp}{\sqrt{\varepsilon_b-1+p^2}-\varepsilon_bp}.\hskip1.8cm\nonumber
\end{eqnarray}
Here, $\varepsilon$ is given by Eq.~(\ref{Drude}) with plasma frequency of Eq.~(\ref{plasmaFy}) and the functions $\varphi$ and $\psi$ represent the metallic-type EM response in the SWCN alignment direction and the dielectric background response in the in-plane perpendicular direction, respectively.

\begin{figure}[t]
\includegraphics[width=0.75\linewidth]{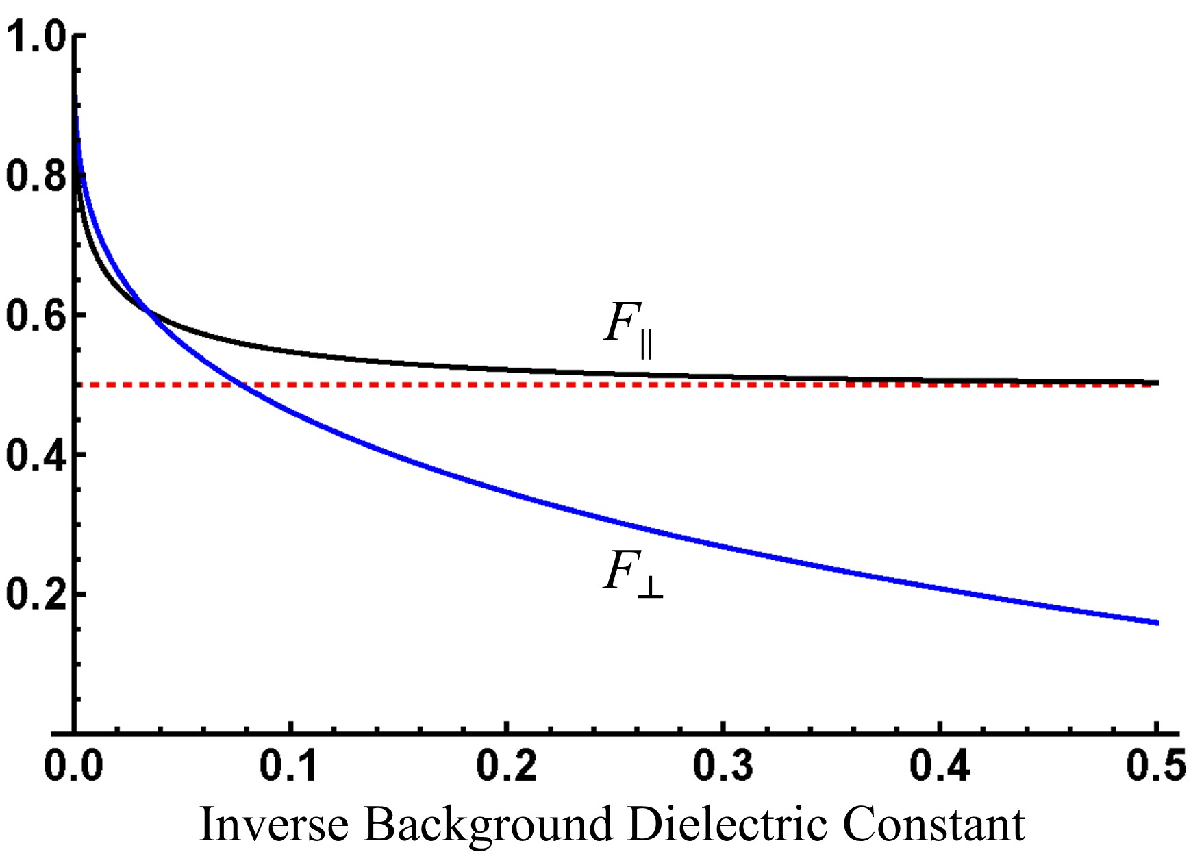}
\caption{Orientational anisotropy of the attractive force for free-standing slabs sketched in Fig.~\ref{fig1}~(b) and (c), relative to the Casimir force.}
\label{fig2}
\end{figure}

Now the procedure described above to derive Eq.~(\ref{NLiso}) can be repeated for $F_{\parallel}$ and $F_{\perp}$ of Eq.~(\ref{Fparper}) separately. This results in the attractive force expressions representative of the SWCN array with parameters $R$, $\Delta$ and slab thickness $d$, as follows
\begin{eqnarray}
F_\parallel=F_C\!\left[\frac{1}{2}+\frac{15}{2\pi^4}\!\int_{0}^{\infty}\!\!\!\!\!\!dx\!\int_1^\infty\!\!\!\!\!\!\!dp\,\frac{x^3}{p^2}\frac{1}{\varphi^2e^x-1}\right.\hskip7.3cm\nonumber\\[-0.1cm]
\label{metmet}\\
\left.-\frac{15c}{\pi^4\omega_p^{3D}l}\!\int_{0}^{\infty}\!\!\!\!\!dx\!\int_1^\infty\!\!\!\!\!\!\!dp\,\frac{x^4}{p^4}\frac{e^x}{\big(e^x-1\big)^{\!2}}\sqrt{\frac{\Delta}{4\pi R}\frac{\alpha(1+R\alpha/\tilde{\varepsilon}d)}{I_0(1/\alpha)K_0(1/\alpha)}}\,\right]\nonumber\\
\mbox{with}~\alpha=\frac{2l}{R}\frac{p}{x\sqrt{p^2-1}}~\mbox{and}\hskip11cm\nonumber\\
F_\perp=F_C\left\{\frac{15}{2\pi^4}\!\int_{0}^{\infty}\!\!\!\!\!\!dx\!\int_1^\infty\!\!\!\!\!\!dp\,\frac{x^3}{p^2}\Big(\frac{1}{\varphi e^x-1}-\frac{1}{\psi e^x+1}\Big)\right.\hskip5.5cm\nonumber\\[-0.1cm]
\label{metdiel}\\
\left.-\frac{15c}{2\pi^4\omega_p^{3D}l}\!\int_{0}^{\infty}\!\!\!\!\!dx\!\int_1^\infty\!\!\!\!\!\!dp\,\frac{x^4e^x}{p^3}\Big[\frac{\varphi\,p}{\big(\varphi e^x-1\big)^{\!2}}-\frac{\psi/p}{\big(\psi e^x+1\big)^{\!2}}\Big]
\sqrt{\frac{\Delta}{4\pi R}\frac{\alpha(1+R\alpha/\tilde{\varepsilon}d)}{I_0(1/\alpha)K_0(1/\alpha)}}\,\right\}.\nonumber
\end{eqnarray}
Taking the limit $\omega_p^{3D}\!\rightarrow\!+\infty$ in these equations is the same as taking the limit $\omega\!\rightarrow\!0$ in Eq.~(\ref{Drude}). This leaves us with the main expansion term alone in both equations, of which that in Eq.~(\ref{metdiel}) reproduces the metal-dielectric attractive force first reported by Lifshitz in his seminal work~\cite{Lifshitz}. Both terms are shown in Fig.~\ref{fig2} as functions of $1/\varepsilon_b$ relative to $F_C$, the Casimir force, to demonstrate the role of the dielectric background in the orientational anisotropy of the attractive force in our system. The anisotropy increases with decreasing $\varepsilon_b$ as expected due to the relative increase of the unidirectional metallic-type EM response component. Both terms become equal to $F_C$ in the limit $\varepsilon_b\!\rightarrow\!\infty$, an analogue of the perfect metal case.

\begin{figure}[t]
\includegraphics[width=1.0\linewidth]{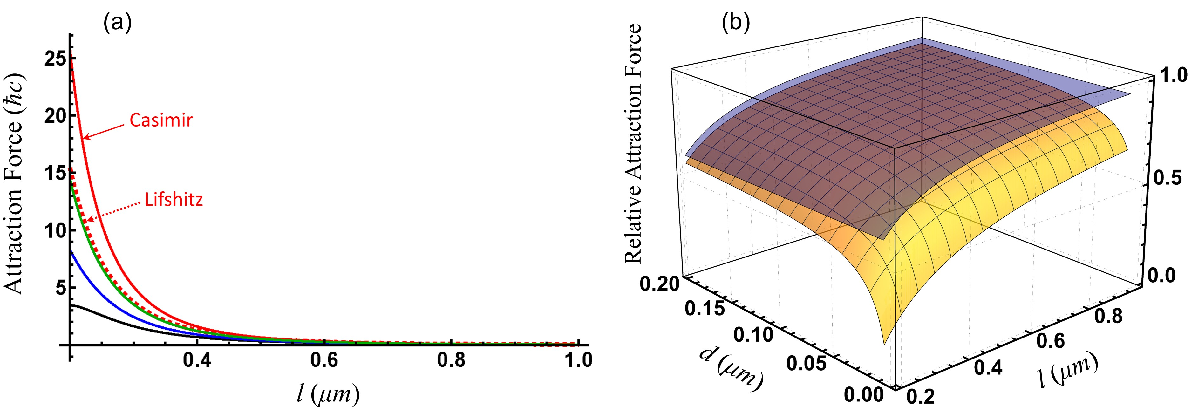}
\caption{(a)~Nonlocal attraction force for a pair of free-standing in-plane isotropic slabs of $10$~nm (black), $20$~nm (blue) and $200$~nm (green) in thickness, as given by Eq.~(\ref{NLiso}). Short-dashed red line is the Lifshitz force of Eq.~(\ref{Lifshitzforce}). (b)~Lifshitz force (dark blue) and nonlocal attraction force (yellow) as functions of the slab thickness and inter-slab distance, relative to the Casimir force.}
\label{fig3}
\end{figure}

\section{Numerical results and discussion}

For all three cases of space-separated finite-thickness slabs shown in Fig.~\ref{fig1}~(a), (b) and (c), the bulk plasma frequency is the main parameter of the long-range nonlocal attractive force in Eqs.~(\ref{NLiso}), (\ref{metmet}) and (\ref{metdiel}) we have obtained. We use a typical free-electron-gas value $\omega_p^{3D}\!=2\times10^{16}$~s$^{-1}$ estimated from $N_{3D}\!\approx\!10^{23}~\mbox{cm}^{-3}$ in our calculations we discuss below.

\subsection{In-plane isotropic TD systems}

Figure~\ref{fig3} shows our numerical results for the nonlocal attraction force obtained from Eq.~(\ref{NLiso}) for the free-standing slabs of varied thickness. The background in-plane dielectric permittivity constant $\varepsilon_b\!=\!9$ was used in these calculations, which is close to those reported experimentally for typical ultrathin plasmonic films~\cite{NL22TiN,BiehsBond2022}. Figure~\ref{fig3}~(a) compares the force calculated for the three slab thicknesses, $d\!=\!10$~nm, $20$~nm and $200$~nm (black, blue and green line, respectively), to the Lifshitz force of Eq.~(\ref{Lifshitzforce}) and to the Casimir force for perfect metals of Eq.~(\ref{Casimir}). It can be seen that the confinement-induced nonlocality weakens significantly the attraction of the thinner slabs. Increasing of the slab thickness diminishes this effect, making the force approach the local Lifshitz limit, which in itself is significantly less than the Casimir force at shorter inter-slab distances. This can also be seen in Fig.~\ref{fig3}~(b), which shows the nonlocal attraction force together with the Lifshitz force relative to the Casimir force as functions of slab thickness and inter-slab distance. Contrary to the Lifshitz force, the nonlocal attraction force can now be seen to quickly drop down with decreasing slab thickness, approaching the form in Eq.~(\ref{NLisothin}) where the nonlocal correction term is $\sim\!1/\sqrt{l}$ as opposed to the $1/l$ dependence of the local Lifshitz force in Eq.~(\ref{Lifshitzforce}).

\begin{figure}[t]
\includegraphics[width=1.0\linewidth]{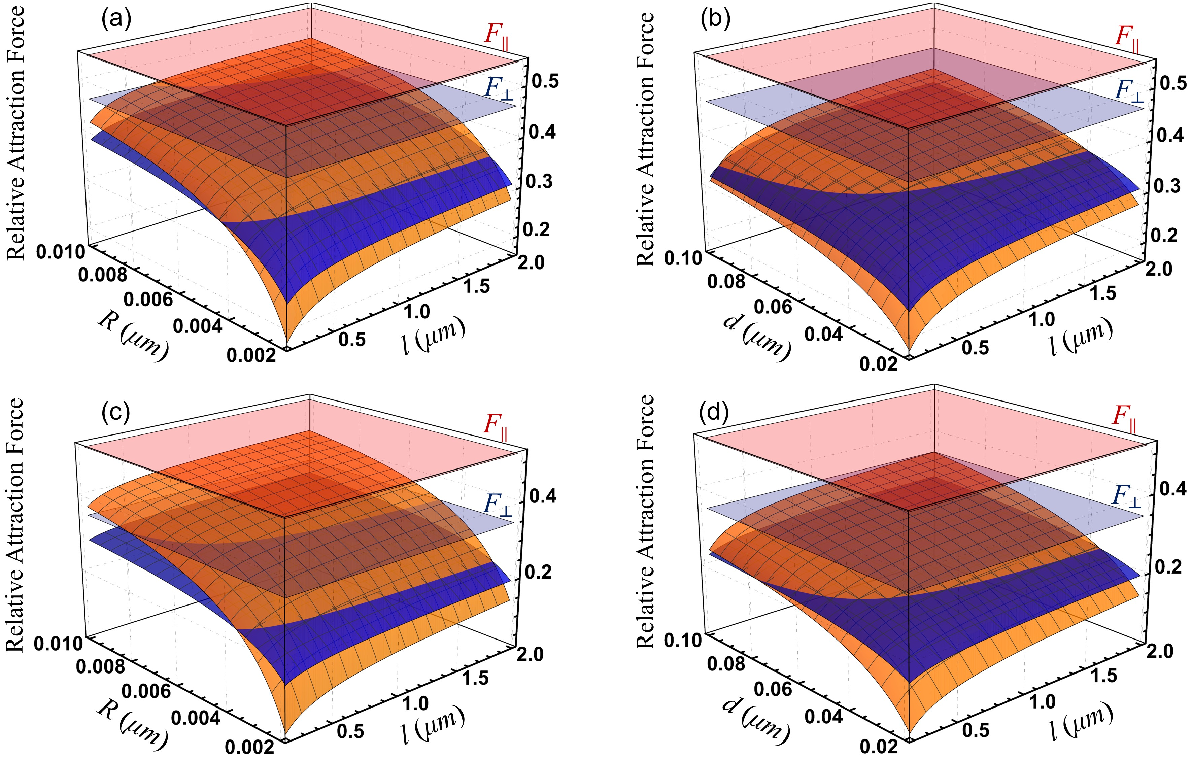}
\caption{Anisotropic nonlocal attraction forces $F_{\parallel}$ and $F_{\perp}$ as functions of the slab thickness and inter-slab distance, relative to the Casimir force. (a)~$5$-monolayer-thick free-standing slabs composed of SWCNs of increasing radius and (b)~slabs of increasing thickness composed of $2$~nm-radius SWCNs, as given by Eqs.~(\ref{metmet}) and (\ref{metdiel}) with $\varepsilon_b\!=\!10$. In (c) and (d) same is shown for $\varepsilon_b\!=\!5$. Lightly colored planes on top show the main expansion terms given by Eqs.~(\ref{metmet}) and (\ref{metdiel}) in the limit $\omega_p^{3D}\!\rightarrow\!+\infty$, or $\omega\!\rightarrow\!0$ in Eq.~(\ref{Drude}), whereby the SWCN array structural parameters are gone.}
\label{fig4}
\end{figure}

\subsection{In-plane anisotropic TD systems}

Figure~\ref{fig4} summarizes our numerical data obtained for the slab thickness and inter-slab distance dependences of the anisotropic nonlocal attraction forces $F_{\parallel}$ and $F_{\perp}$ (relative to the Casimir force) as given by Eqs.~(\ref{metmet}) and (\ref{metdiel}) for densely packed ($\Delta\!=\!2R$) free-standing SWCN arrays oriented as shown in Fig.~\ref{fig1}~(b) and (c), respectively.~In (a), the forces are shown for the five-monolayer-thick slabs composed of SWCNs of increasing radius.~In (b), they are shown for the slabs composed of the $2$~nm-radius SWCN array monolayer of increasing monolayer number. In both cases the slab thickness increases either due to the SWCN radius with monolayer number fixed, or due to the monolayer number with SWCN radius fixed. Lightly colored planes on top show the main expansion terms of Eqs.~(\ref{metmet}) and (\ref{metdiel}) one obtains in the limit $\omega_p^{3D}\!\rightarrow\!+\infty$, or $\omega\!\rightarrow\!0$ in Eq.~(\ref{Drude}), whereby the SWCN array structural parameters disappear and the dielectric background constant $\varepsilon_b$ remains the only parameter to control the difference between $F_{\parallel}$ and $F_{\perp}$ as discussed for Fig.~\ref{fig2}. Both (a) and (b) show the graphs calculated for $\varepsilon_b\!=\!10$ (cf. Fig.~\ref{fig2}). In (c) and (d) the same is shown for $\varepsilon_b\!=\!5$.

In addition to the reduction of the attractive force with decreasing slab thickness we have seen above for the in-plane isotropic case, there is one more remarkable non-obvious feature that can now be seen in Fig.~\ref{fig4}. This is the crossover behavior of the attractive force: namely, the force $F_{\parallel}$, which is always greater than $F_\perp$ for thicker slabs, weakens quickly with slab thickness reduction and becomes less than the force $F_\perp$ for sufficiently thin slabs. This means that a pair of ultrathin slabs prefers to stick together in the perpendicularly oriented manner sketched in Fig.~\ref{fig1}~(c), counter-intuitively, rather than in the parallel relative orientation shown in Fig.~\ref{fig1}~(b) one would think of customarily. Although not obvious at first glance, this crossover behavior can still be understood as being due to the increased transmission of the metallic-type EM response component in our system, facilitated by the plasma frequency red shift as the slab thickness $d$ in Eq.~(\ref{plasmaFy}) decreases. Metals are known to be transmissive in the frequency range above their plasma frequency (see, e.g., Ref.\cite{Kittel}), which now shifts towards infrared. Increased metallic transmission reduces photon absorption necessary for inter-slab attraction to occur~\cite{Lifshitz,LandauLifshitz}. Hence, it follows from Eq.~(\ref{Fparper}) that a pair of parallel oriented slabs with metal-metal/dielectric-dielectric type of virtual photon exchange becomes less attractive than a pair of perpendicularly oriented slabs with metal-dielectric/dielectric-metal type of virtual photon exchange as the former is associated with twice less absorption (roughly) than the latter. Decreasing of the dielectric background constant $\varepsilon_b$ leads to the relative weight increase for the metallic type EM response, just like it would be a pair of thicker slabs, which is why the crossover effect in (c) and (d) can be seen to shift toward smaller slab thicknesses as compared to those in (a) and (b).

\subsection{Can the Lifshitz formula be applied to ultrathin films ?}

The Lifshitz formula we use herein was originally obtained for two media filling space-separated half-spaces with plane-parallel boundaries~\cite{Lifshitz}. In practice, however, one always deals with finite-thickness material slabs. Hence, there are two questions we should answer to justify the Lifshitz formula applicability in our case: (i)~is it appropriate to replace bulk material EM response functions of the Lifshitz model by their thickness dependent in-plane counterparts we use, and (ii)~what is the smallest thickness for a material slab to correctly represent medium-filled half-space. Both questions are addressed in what follows.

For question (i), we note that to obtain the attraction force between the two space-separated material half-spaces as a function of their out-of-plane separation distance alone, one necessarily has to integrate over the two in-plane directions. This is what the Lifshitz model does assuming (customarily) the bulk-medium linear EM response to be the same in all three directions, of which two in-plane directions are integrated over while the out-of-plane one remains to represent the distance dependence of interest. This is equivalent to using the thickness-dependent in-plane EM response functions in the integration directions we do here, whereby the Lifshitz model consistency is preserved.

In order to answer question (ii), it is sufficient to compare the $s$ and $p$ wave reflection coefficients $r_{s,p}$ given by the reciprocals of the pre-exponential factors in the first and second terms of Eq.~(\ref{Lifshitz}), respectively, with those of the free-standing film of thickness $d$ (see, e.g., Refs.\cite{BiehsBond2022,BondPRR20})
\[
R_{s,p}=\frac{r_{s,p}\Big[1-e^{2id\sqrt{\varepsilon(k,\omega)(\omega/c)^2-k^{2}}}\,\Big]}{1-r_{s,p}^2\,e^{2id\sqrt{\varepsilon(k,\omega)(\omega/c)^2-k^{2}}}}
\]
after the required substitutions of variables $\omega p/c\!=\!\sqrt{(\omega/c)^2-k^2}$, $\omega\!=\!ixc/(2pl)$ are done. Here, the exponential factor is responsible for the backscattering from the second interface that is now present in the system. Using Eq.~(\ref{Drude}) with damping neglected as before, we have
\[
2id\sqrt{\varepsilon(k,\omega)\frac{\omega^2}{c^2}-k^{2}}=-2d\frac{\omega_p^{3D}}{c}\sqrt{1+\frac{x^2(p^2+\varepsilon_b-1)}{p^2}\!\left(\!\frac{c}{2l\omega_p^{3D}}\!\right)^{\!\!2}},
\]
where $x\!\sim\!1$ and $p\!\ge\!1$ as it follows from the structure of the integral expression in Eq.~(\ref{Lifshitz}). We also have $c/\omega_p^{3D}\!=3\!\times\!10^8\,\mbox{m\,s}^{-1}/(2\!\times\!10^{16}\,\mbox{s}^{-1})\!=\!15$~nm, so that $c/(2l\omega_p^{3D})\!<\!1$ for all $l\!>\!10$~nm and $2d\omega_p^{3D}/c\!>\!1$ for all $d\!>\!10$~nm, whereby the backscattering from the second interface becomes exponentially suppressed making the coefficients $R_{s,p}$ indistinguishable from $r_{s,p}$ for films greater than $10$~nm in thickness. The data we present in Figs.~\ref{fig3}~and~4 above never fall out of this range.

\section{Conclusion}

In this work, we use the Lifshitz theory to explore the impact of the confinement-induced nonlocality of the material EM response on the (Casimir) force of attraction between two space-separated TD material slabs. The Lifshitz theory represents molecular attractive forces between solids in terms of their respective linear EM response functions for which we use those provided by the confinement-induced nonlocal EM response model~\cite{BondOMEX17,BondOMEX19}. Confinement-induced is a special type of nonlocality that comes from the KR pairwise electron interaction potential in optically dense ultrathin films of finite thickness~\cite{KRK,KRR}. This potential is stronger than the Coulomb interaction potential and depends on the film thickness, which is why our EM response model covers both ultrathin ($d\!\gtrsim\!10\,$nm) and conventional thin films~\cite{BiehsBond2022}. In general, the Lindhard-Mermin nonlocality of bulk materials might also play a role for thicker slabs~\cite{ChapuisEtAl2008}; however, this is the $k^2$-infinitesimal order nonlocality, which in the most important low-momentum domain is much less than the confinement-induced $k$-infinitesimal order nonlocality of the KR model we use~\cite{BondOMEX17}.

We calculate the long-range attractive forces for in-plane isotropic and anisotropic free-standing TD material slabs. In the former case, we show that the confinement-induced nonlocality not only weakens the attraction of ultrathin slabs but also changes the distance dependence of the material-dependent correction to the Casimir force to go as $\sim\!1/\sqrt{l}$ contrary to the $\sim\!1/l$ dependence of that of the local Lifshitz force. Increasing the slab thickness diminishes this effect, making the force approach the local Lifshitz force limit which in itself is significantly less than the material-independent Casimir force. In the latter case, we use closely packed array of single-diameter parallel aligned SWCNs in a dielectric layer of finite thickness to demonstrate strong orientational anisotropy and crossover behavior for the inter-slab attractive force in addition to its reduction with decreasing slab thickness. We show that and we explain why a pair of ultrathin slabs prefers to stick together in the perpendicularly oriented manner, rather than in the parallel relative orientation as one would customarily expect and as it is shown to occur for thicker slabs.

Our results are obtained for optically dense TD films in the long-range low-temperature quantum limit $lk_BT/(\hbar c)\!\ll\!1$ of the Lifshitz theory~\cite{Lifshitz}. For $l\!\lesssim\!1\,\mu$m where the effects we discuss occur, this becomes $T\!\ll\!2000$~K so that high-temperature calculations in the classical limit will hardly affect our reported results. Another case is that of closely-separated diluted quasi-2D films, where the electron band-structure and interactions are predominantly controlled by the dimensionality, and so temperature effects as well as purely quantum effects such as electron energy-level quantization and wavefunction spill-out~\cite{BenassiCalandra}, which are not involved in our case, can be important~\cite{LePabLil2022}.

\section{Acknowledgements}

I.V.B. is supported by the U.S. Army Research Office under award No. W911NF2310206. M.D.P. was supported by the U.S. National Science Foundation grant No.~DMR-1830874 (awarded to I.V.B.). P.R.-L. acknowledges support from AYUDA PUENTE 2022, URJC and QuantUM program of the University of Montpellier. P. R.-L. would like to thank the University of Montpellier and the theory group for Light-Matter and Quantum Phenomena of the Laboratoire Charles Coulomb for hospitality during his stay in Montpellier where part of this work was done. L.M.W. acknowledges financial support from the US Department of Energy under Grant No. DE-FG02-06ER46297. M.A. acknowledges the grant "CAT", No.~A-HKUST604/20, from the ANR/RGC Joint Research Scheme sponsored by the French National Research Agency (ANR) and the Research Grants Council (RGC) of the Hong Kong Special Administrative Region. I.V.B., L.M.W. and M.A. gratefully acknowledge support from the Kavli Institute for Theoretical Physics (KITP), UC Santa Barbara, under U.S. National Science Foundation Grant No.~PHY-1748958, where this collaborative work was started. I.V.B. acknowledges KITP hospitality during his invited visit as a KITP Fellow 2022--23 made possible by the Heising-Simons Foundation.


\end{document}